\documentclass[12pt,preprint]{aastex}

\shorttitle{Dust Formation in Cas A}
\shortauthors{}

\newcommand{\etal}{{et al.\/~}}

\newcommand{\spitzer}{\textit{Spitzer}}
\def\lowermass{0.020}
\def\uppermass{0.054}

\begin{document}
\title{Freshly Formed Dust in the Cassiopeia A Supernova Remnant
as Revealed by the {\it Spitzer} Space Telescope}

\author{
J. Rho\altaffilmark{1}, T. Kozasa\altaffilmark{2}, W. T. Reach\altaffilmark{1},
J. D. Smith\altaffilmark{3}, L. Rudnick\altaffilmark{4}, T. DeLaney\altaffilmark{5}, 
J. A. Ennis\altaffilmark{4}, H. Gomez\altaffilmark{6}, A. Tappe\altaffilmark{1,7}
}
\altaffiltext{1}{{\it Spitzer} Science Center, California
Institute of Technology,  Pasadena, CA 91125; rho, reach@ipac.caltech.edu} 
\altaffiltext{2}{Department of Cosmosciences, Graduate School of Science,  Hokkaido University, 
Sapporo 060-0810, Japan}
\altaffiltext{3}{Steward Observatory,
933 N. Cherry Ave., Tucson, AZ 85712}
\altaffiltext{4}{Department of Astronomy, University of Minnesota, 116 Church St. SE, Minneapolis, MN 55455}  
\altaffiltext{5}{MIT Kavli Institute, 77 Massachusetts Ave., Room NE80-6079,
Cambridge, MA 02139}
\altaffiltext{6}{School of Physics and Astronomy, University of Wales,
Cardiff, Wales, UK} 
\altaffiltext{7}{Harvard-Smithsonian Center for Astrophysics, 60 Garden Street, Cambridge, MA 02138}

\begin{abstract}

We performed {\it Spitzer} Infrared Spectrograph mapping observations 
covering nearly the entire extent of the Cassiopeia A supernova remnant
(SNR), producing mid-infrared (5.5-35 $\mu$m) spectra every
5$\arcsec$-10$\arcsec$.   Gas lines of Ar, Ne, O, Si, S and Fe, and
dust continua were strong for most  positions. We identify three
distinct ejecta dust  populations based on their continuum shapes.  The
dominant dust continuum  shape exhibits a strong peak at 21 $\mu$m.  A
line-free map of 21 $\mu$m-peak  dust made from the 19-23 $\mu$m range
closely  resembles the [Ar~II], [O~IV], and [Ne~II] ejecta-line maps
implying that  dust is freshly formed in the ejecta.  Spectral fitting
implies the  presence of SiO$_2$, Mg protosilicates, and FeO grains in
these regions.   The  second dust type exhibits a rising continuum up
to 21 $\mu$m and then flattens  thereafter.  This ``weak 21 $\mu$m''
dust is likely composed of Al$_2$O$_3$ and C grains.  The third dust
continuum shape is featureless with a gently  rising spectrum and is
likely composed of MgSiO$_3$ and either Al$_2$O$_3$ or Fe grains.  
Using the least massive composition  for each of the three dust classes
yields a total mass of \lowermass\ M$_\odot$. Using the most-massive
composition yields a total mass of \uppermass\  M$_\odot$.  The primary
uncertainty in the total  dust mass stems from the selection of the
dust composition  necessary for fitting  the featureless dust as well
as 70 $\mu$m flux.    The freshly formed dust mass derived from Cas A 
is sufficient from SNe to explain the lower limit on the dust masses in
high redshift galaxies.

\end{abstract}

\section{Introduction}

The recent discovery of huge quantities of dust
($10^8-10^9\,\rm{M_{\odot}}$)  in very high--redshifted galaxies and
quasars (Isaak et al. 2002; Bertoldi et al. 2003) suggests that  dust
was produced efficiently in the first generation of supernovae (SNe).
Theoretical studies (Kozasa et al. 1991; Todini \& Ferrara 2001, 
hereafter TF; Nozawa etal. 2003, N03) predicted the formation of a
significant quantity of dust ($\sim 0.1-1.0$ $M_{\odot}$)  in the
ejecta of Type II SNe, and the predicted dust  mass is believed to be
sufficient to  account for the quantity of dust observed at high
redshifts  (Maiolino et al. 2006; Meikle et al. 2007).  Recently, a
model of dust evolution in high--redshift galaxies (Dwek et al. 2007)
indicates that at least 1 $M_{\odot}$ of dust per SN is necessary for
reproducing the observed dust mass in one hyperluminous quasar at
$z=6.4$. Observationally,  the presence of freshly formed dust has been
confirmed in a few core-collapsed SNe, such as SN1987A, which clearly
have showed several signs of  dust formation in the ejecta (see McCray
1993 for details).    The highest dust mass obtained so far for SN
1987A is $7.5 \times 10^{-4}$ $M_{\odot}$ \citep{erc07}.   Recent {\it
Spitzer} and  {\it HST} observations (Sugerman et al. 2006) showed 
that up to 0.02 $M_{\odot}$ of dust  formed in the ejecta of SN2003gd
with the progenitor mass of   6--12 $M_{\odot}$,  and the authors
concluded  that SNe are major dust factories. However, from the
detailed analysis of the  late--time mid--infrared observations, 
Meikle et al. (2007) found  that the mass of freshly formed dust in the
same SN is only $4 \times 10^{-5} M_{\odot}$, and failed to confirm the
presence 0.02 $M_{\odot}$  dust in the ejecta. The aforementioned
results   show that the derived dust mass is model-dependent, and  
that the amount of dust that really condenses in the ejecta of
core-collapsed SNe is  unknown.

Cassiopeia A  (Cas A) is the only Galactic supernova remnant (SNR) that
exhibits  clear evidence of  dust formed in ejecta (Lagage \etal 1996;
Arendt \etal 1999,  hereafter ADM). The amount of dust that forms in
the ejecta of young SNR is still controversial.    Previous
observations inferred only $<3\times 10^{-3}\, \rm{M_{\odot}}$ of dust 
at temperatures between 90 and 350 K  (ADM; Douvion et al. 2001,
hereafter D01) ; this estimate is 2 to 3 orders of magnitude too little
to explain the dust    observed in the early Universe.   Recent
submillimeter observations of Cas\,A and Kepler with SCUBA
\citep{dun03, mor03a} revealed the presence of large amounts of cold
dust ($\sim    0.3-2\,\rm{M_{\odot}}$ at 15--20 K) missed by previous
IRAS/ISO    observations.   On the other hand,   highly elongated
conductive needles with mass of  only 10$^{-4}$ to 10$^{-3}$
$M_{\odot}$  could also explain a high sub-mm flux of Cas A, when
including grain destruction by sputtering (Dwek 2004),  though the
physicality of such needles is doubtful \citep{gom05}.  While
\cite{kra04} showed that much of the 160$\mu$m emission observed with
Multiband Imaging Photometer for {\it Spitzer} (MIPS) is foreground
material, suggesting there is no cold dust in Cas A, \cite{wil05}  used
CO emission towards the remnant   to show that up to about a solar mass
of dust could still be associated   with the ejecta, not with the
foreground material.  These controversial scenarios  of dust mass 
highlight the importance of correctly identifying the features  and
masses of dust freshly formed in  Cas A.  The Galactic young SNR Cas A
allows us to study in detail the distribution and the  compositions of
the dust relative to the ejecta and forward shock with infrared
Spectrograph onboard the {\it Spitzer Space Telescope}.  

Cas A is one of the youngest Galactic SNRs with an age of 335 yr
attributed to a SN explosion in AD 1671. The progenitor of Cas A is
believed to be a Wolf-Rayet star with high nitrogen abundance
\citep{fes01} and to have a mass of 15-25 M$_{\odot}$
\citep{kif01,you06} or 29-30 M$_{\odot}$ \citep{per02}.   The predicted
dust mass formed in SNe depends on the progenitor mass; for  a
progenitor mass of 15 to 30 M$_{\odot}$, the predicted dust mass is
from 0.3 to 1.1 M$_{\odot}$ (NO3) and  from 0.08 to 1.0 M$_{\odot}$ 
(TF), respectively.  In this paper, we present {\it Spitzer} Infrared
Spectrograph (IRS) mapping  observations of Cas A, and identify three
distinct classes of dust associated  with the ejecta and discuss dust
formation and composition with an estimate of the total mass of freshly
formed dust.


\section{The IRS Spectra and Dust Maps} We performed {\it Spitzer} IRS
mapping observations covering nearly the entire extent of Cas A on 2005
January 13 with a total exposure time of 11.3 hr.  The  Short Low (SL:
5-15 $\mu$m) and Long Low (LL: 15-40 $\mu$m) IRS mapping  involved
$\sim$16$\times 360$ and $4\times 91$ pointings, producing spectra 
every 5$\arcsec$ and 10$\arcsec$, respectively.  The spectra were
processed  with the S12 version of the IRS pipeline using the CUBISM
package  (Kennicutt et al. 2003;  Smith et al. 2007), whereby
backgrounds were  subtracted and an extended emission correction was
applied.  The spectral  resolving power of the IRS SL and LL modules
ranges from 62 to 124.  

\begin{figure}
\epsscale{0.9}
\plotone{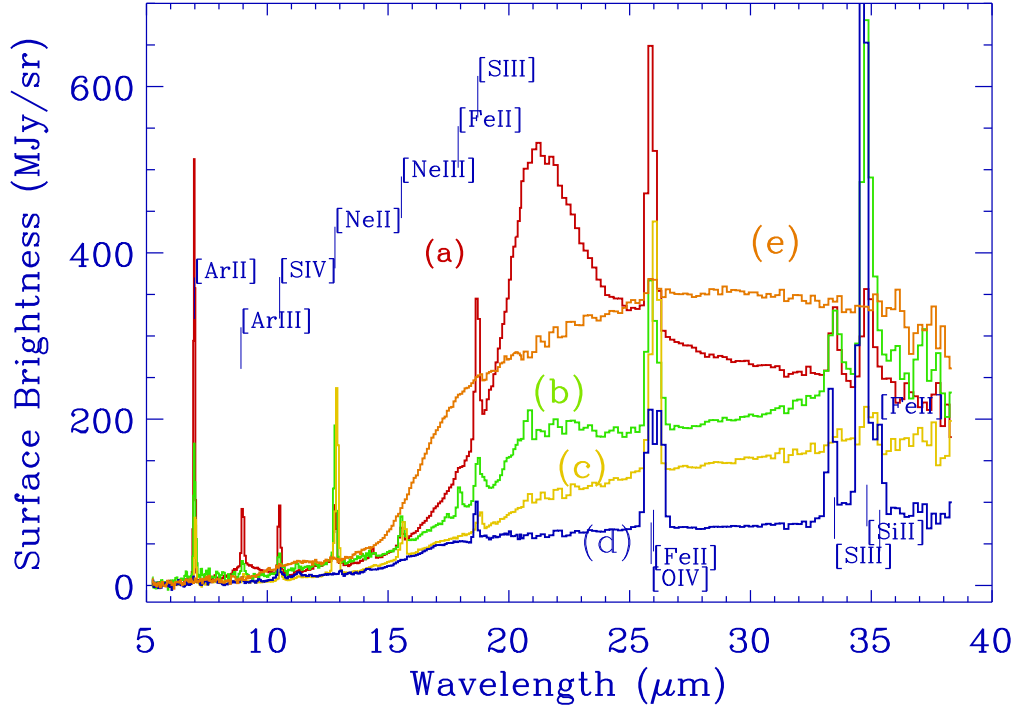}
\caption{Representative Set of {\it Spitzer} IRS spectra of Cas A. The
contrast between continuum shape and line emission is noticeable.  Many
of the strong ejecta lines are seen including [Ar~II] (7 $\mu$m), [Ar~
III] (8.9 $\mu$m), [Si~IV] (10.5 $\mu$m), [Ne~ II] (12.8 $\mu$m), [Ne~
III] (15.5 $\mu$m),  [S~III](18.7 $\mu$m),  [O~IV] and [Fe~II]
(26 $\mu$m), [S~III] (33.5 $\mu$m), [Si~II] (34.8 $\mu$m), and [Fe~ II]
(35.3 $\mu$m).  The dominant continuum shapes are 21 $\mu$m-peak dust
showing a dust feature at 21 $\mu$m often accompanied by a silicate
emission feature at 9.8 $\mu$m with strong Ar lines (red, curve a) and
weak-21 $\mu$m dust with relatively strong Ne lines compared with Ar (green curve, b). Featureless
spectra include the continuous rising spectra (yellow, c),  and the
gently rising spectra (blue, d) with strong O+Fe or Si lines. 
 The positions of RA. and Dec. are 350.900, 58.8356 (a), 350.812, 58.8075 (b), and 
 350.879    58.7911 (c), 350.857,
58.815 (d), and 350.862    58.8550 9 (e). 
For illustration, the spectra were multiplied by 1.4 (curve a), 2.8 (b), 0.8 (c),  
1.5 (d), and 2 (e), respectively.
``Broad" continuum spectra arise from interstellar/circumstellar medium (orange, e). 
}
\label{sixspec}
\end{figure}

The IRS spectra of Cas A show bright ejecta emission lines from Ar, Ne,
S,  Si, O, and Fe and various continuum shapes as indicated by the
representative  spectra in Figure 1.  The most common continuum shape
exhibits a large bump  peaking at 21 $\mu$m as shown by spectrum ``a''
in Figure \ref{sixspec}.   This ``21 $\mu$m-peak'' dust is often
accompanied by the silicate emission  feature at 9.8 $\mu$m which
corresponds to the stretching mode.  A second  class of continuum
shapes exhibits a rather sharp rise up to 21 $\mu$m and  then stays
flat thereafter.  This ``weak-21 $\mu$m dust'' is often associated 
with relatively strong Ne lines (in comparison with Ar lines) and is
indicated by spectrum ``b'' in Figure \ref{sixspec}.  The third type of
dust continuum is characterized by a smooth  and featureless, gently
rising spectrum with strong  [\ion{O}{4}]+[\ion{Fe}{2}] and
[\ion{Si}{2}] emission lines as shown by  spectra ``c'' and ``d'' in
Figure \ref{sixspec}.  The spectrum ``d'' shows  double line structures
that may be due to doppler-resolved lines of [\ion{O}{4}]  at 26 $\mu$m
and [\ion{Si}{2}]  at 35 $\mu$m. Note that the ``featureless'' dust
(spectrum ``d'' in Fig. \ref{sixspec}) is a  class of dust, separate
from the interstellar/circumstellar dust    (spectrum ``e'' in Fig.
\ref{sixspec})  heated by the forward shock.  The
interstellar/circumstellar dust spectrum in Cas A has no associated gas
line emission. The  ``broad'' continuum   (see Figure 7b of Ennis et
al. 2006) is a combination of  the spectra ``c'' and ``e''.  The
spectrum ``c''  has contamination from the shock heated dust in
projection,  and for simplicity it is excluded in estimating the masses
of the freshly formed dust   (see \S5).  The ``featureless'' dust lacks
the gentle peak around  26 $\mu$m and also lacks the interstellar
silicate-emission feature between  9 $\mu$m and 11 $\mu$m observed in
the spectra from the forward shock region. Most importantly,  the
``featureless'' dust accompanies relatively strong Si and S ejecta
lines and mostly from the interior of the remnant (blue region in Fig.
\ref{images}f). 

We generated a map of the 21 $\mu$m-peak dust by summing the emission
over 19-23 $\mu$m after subtracting a baseline between 18-19 $\mu$m and
23-24 $\mu$m.   The line-free dust map (Fig. ~\ref{images}a) resembles 
the [Ar~II] and [O~IV]+[Fe~II]  ejecta-line maps, as shown in Figures
~\ref{images}b and  ~\ref{images}c,  and we also find that the [Ne~II]
map is very similar to the [Ar~II] map. The [\ion{Ar}{2}] map shows a 
remarkable similarity to the 21 $\mu$m-peak dust map (Fig.
\ref{images}a and  \ref{images}b),  thereby confirming this  dust is
freshly formed in the ejecta.  Maps of [\ion{Si}{2}] (Fig.
\ref{images}d) and [O~IV]+[Fe~II] (Fig. ~\ref{images}c) shows
significant  emission at the center revealing ejecta that have not yet
been overrun by the  reverse shock (unshocked ejecta).   There is also
[\ion{Si}{2}] and  [O~IV]+[Fe~II] emission at the bright ring
indicating that some of the Si and O+Fe ejecta have recently 
encountered the reverse shock. While the bright O+Fe emission outlines
the same bright ring structure as the [Ar~II] and 21 $\mu$m-peak dust
maps, the bright part of the Si shell shows a different morphology from
the other ejecta maps.

We can characterize the spectra of our three dust classes by using the flux 
ratios between 17 $\mu$m and 21 $\mu$m and between 21 $\mu$m and 
24 $\mu$m.  Although the spectra in Cas A show continuous changes in
continuum shape from strong 21 $\mu$m peak to weak 21 $\mu$m peak and to 
featureless, we can locate regions where each of the three classes dominates.  
Figure \ref{images}f shows the spatial distribution of our three dust classes 
where red, green, and blue indicate 21 $\mu$m-peak dust, weak-21 $\mu$m dust, 
and featureless dust, respectively.  The flux ratios used to identify the 
three dust classes are as follows where $I_\lambda$ is the flux
density in the extracted spectrum at wavelength $\lambda$ ($\mu$m):

1) 21 $\mu$m-peak dust: we use the ratio $I_{21}/I_{24} > 1+\sigma_{21/24}$,
where $\sigma_{21/24}$ is the dispersion in $I_{21}/I_{24}$ over the remnant, which
is equivalent to $I_{21}/I_{17} \geq 3.4$.  The regions
with 21 $\mu$m-peak dust coincide with the brightest ejecta.

2) Weak-21 $\mu$m dust: we use the ratio $1-\sigma_{21/24} < I_{21}/I_{24} < 
1 + \sigma_{21/24}$, which is equivalent to $2.3 <I_{21}/I_{17} < 3.4$.  The 
regions showing the weak-21 $\mu$m continuum shape mostly coincide with faint 
ejecta emission, but not always.

3) Featureless dust map: we use the ratio $I_{21}/I_{24}< 1 - \sigma_{21/24}$,
which is equivalent to $I_{21}/I_{17} < 2.3$.  This ratio also picks
out  circumstellar dust heated by the forward shock, so we used several
methods to  exclude and mitigate contamination from circumstellar dust
emission.  First,  using X-ray and radio maps, we excluded the forward
shock regions at the edge  of the radio plateau \citep{got01}.  Second,
there are highly structured  ``continuum-dominated'' X-ray filaments
across the face of the remnant  which are similar to the exterior
forward shock filaments and may be  projected forward shock emission
\citep{del04}.  For our analysis, we  excluded regions where there were
infrared counterparts to the projected  forward shock filaments.  Third,
for simplicity we excluded regions with gently rising  spectra
identified by curve ``c'' (the spectra which continues to rise to longer
wavelengths) in Figure \ref{sixspec}.  This type of spectrum is mainly
found on the eastern side of Cas A where there is an H$\alpha$  region,
the northeast jet, and other exterior optical ejecta \citep{fes01} 
making it difficult to determine if the continuum emission is due to
ejecta  dust or circumstellar dust. However, note that some portion of
the continuum in the spectra, ``c'', is freshly formed dust.   We finally
excluded regions where there was a  noticeable correlation to optical
quasi-stationary flocculi \citep{van71} which are dense  circumstellar knots from the
progenitor wind.  

The featureless dust emission appears primarily across the center of
the remnant, as shown in Figure \ref{images}d (blue). The featureless
dust is accompanied by relatively strong [\ion{Si}{2}] and [\ion{S}{3}]
and [\ion{O}{4}+\ion{Fe}{2}] lines, as shown by the spectrum ``d'' of
Figure \ref{sixspec}.  The  [\ion{O}{4}]+[\ion{Fe}{2}] line map (Fig.
~\ref{images}c)  shows significant emission at  the center as well as
at the bright ring of the reverse shocked material.  The [\ion{Si}{2}]
line map shows different morphology than other line maps and the 21
$\mu$m-peak dust map; depicting center-filled emission with a partial
shell, as shown in Figure \ref{images}.   This poses the following
important question: why is the Si   map   more center-filled than the
Ar map? The answer is unclear because Si and Ar are both expected at
similar depths in the nucleosynthetic layer (e.g. Woosley, Heger \&
Weaver 2002). The relatively faint infrared emission of Si and S at the
reverse shock  may imply relatively less Si and S  in the reverse
shock.  We suspect it is because the Si and S have condensed to  solid
form such as Mg protosilicate, MgSiO$_3$, Mg$_2$SiO$_4$ and FeS.  In
contrast,  Ar remains always in the gas and does not condense to dust,
so it should be infrared or X-ray emitting gas.  Alternate explanation 
is  that the ionization in the interior is due to photoionization from
the X-ray shell (see  Hamilton \& Fesen 1988); in this case, the lack
of interior \ion{Ar}{2} relative to \ion{Si}{2} might be due to its
much higher ionization potential (16 eV compared to 8 eV). Theoretical
models of nucleosynthesis, accounting for heating, photoionization, and
column density of each element would be helpful for understanding the
distribution of nucleosynthetic elements.

The Si and S emission detected at the interior, is most likely
unshocked ejecta where the revere shock has not yet overtaken the ejecta.
The radial profile of unshocked ejecta is centrally peaked at the time
of explosion, as shown by \cite{che89}. The radial profile of unshocked
Fe ejecta is also expected to be  center-filled for $\sim$1000 yr old
Type Ia SNR of SN 1006 \citep{ham88}.   The morphology of the
featureless dust resembles that of unshocked ejecta, supporting the
conclusion that the featureless dust  is also freshly formed dust. The
spectrum in Figure \ref{sixspec} (curve ``d''), shows the resolved two
lines at 26 $\mu$m  and at 35 $\mu$m.  The two respective lines at
$\sim$26 $\mu$m  may be resolved lines of [\ion{O}{4}] and
[\ion{Fe}{2}], and at $\sim$35 $\mu$m [\ion{Si}{2}] and [\ion{Fe}{2}]
(as expected that the unshocked ejecta near the explosion center have a
low velocity); alternatively, they could be highly doppler-shifted
lines (in this case the two lines at 26 $\mu$m are both [\ion{O}{4}],
and the two lines at 35 $\mu$m are both  [\ion{Si}{2}]).  The newly
revealed unshocked ejecta deserves extensive studies; preliminary
doppler-shifted maps were  presented in \cite{del06} and the detailed
analysis of velocities and abundances of unshocked and shocked ejecta
will be presented in future papers \citep{del07, smi07}.

\section{Spectral Fitting and Dust Composition}

We performed spectral fitting to the IRS continua using our example
regions  in Figure \ref{sixspec}.  Included in the fitting are  MIPS 24
$\mu$m and 70 $\mu$m  fluxes \citep{hin04}, and the contribution of
synchrotron emission (Figs. \ref{21umpeakspec} and \ref{weak21umspec}),
estimated from the radio fluxes \citep{del04} and Infrared Array Camera
(IRAC) 3.6 $\mu$m fluxes \citep{enn06}. We measured synchrotron
radiation components for each position using   radio maps and assuming
the spectral index $\alpha$=-0.71 \citep{rho03} where log S $\propto$
$\alpha$ log $\nu$.  Because the full-width-half-maximum of 24 $\mu$m
is  smaller than the IRS extracted region, the surface brightnesses for
24 $\mu$m  were measured using a 15$\arcsec$ box, the same size as the
area used for  the extracted IRS LL spectra.  We also made color
corrections to each MIPS  24 $\mu$m data point based on each IRS
spectrum and band-filter shape; the  correction was as high as 25\% for
some positions.  While the uncertainty of calibration errors in IRAC is
3-4\%, that of MIPS 24 $\mu$m is better than 10\%.   The MIPS 70 $\mu$m
image \citep{hin04}, shown in Figure \ref{images}e, clearly resolves
Cas A from background emission, unlike the 160 $\mu$m image
\citep{kra04}.  Most of the bright 70 $\mu$m emission appears at the
bright ring and corresponds to the 21 $\mu$m dust map and the shocked
ejecta, particularly [\ion{Ar}{2}], indicating that the 70 $\mu$m
emission is primarily from freshly formed dust in the ejecta.  The 70
$\mu$m emission also appears at the interior as shown in Figure
\ref{images}e.   We measured the brightness for 70 $\mu$m within a
circle of radius 20$\arcsec$ for each position, accounting for the
point-spread function (note that when the emission is uniform, the
aperture size does not affect the surface  brightness).  We estimated
the uncertainties of the 70 $\mu$m fluxes to be as large as 30\%.  The
largest uncertainty comes from background variation due to cirrus
structures based on our selection of two background areas,  5$\arcmin$
to the northwest and south of the Cas A. 

The dust continuum is fit with the Planck function (B$_{\nu}(T)$)
multiplied by the absorption efficiency ($Q_{abs}$) for various dust
compositions, varying the amplitude and temperature of each component. 
To determine the dust composition, we consider not only the grain
species predicted by the model of dust formation in SNe (TF, N03), but
also Mg protosilicates (ADM) and FeO \citep{hen95} as possible
contributors to the 21 $\mu$m feature. The  optical constants of the
grain species used in the calculation are the same as those of  \cite{hir05},
except for amorphous Si \citep{pil85}, amorphous SiO$_2$ \citep{phi85},
amorphous Al$_2$O$_3$ \citep{beg97}, FeO \citep{hen95},  and we apply
Mie theory \citep{boh83} to calculate the absorption efficiencies,
Q$_{abs}$,  assuming the grains are spheres of radii $a=0.01$ $\mu$m. 
We fit both amorphous and crystalline grains for each composition, but
it turned out that the fit results in Cas A (see \S3) favor amorphous
over crystalline grains. Thus, default grain composition indicates
amorphous, hereafter.   For Mg protosilicate, the absorption
coefficients are evaluated from the mass absorption  coefficients
tabulated in  \cite{dor80}, and we assume that the absorption
coefficient varies as $\lambda^{-2}$ for $\lambda >$ 40 $\mu$m, typical
for silicates.   We fit the flux density for each spectral type using
scale factors $C_i$ for each grain type $i$, such that F$_{\nu}^i$ =
$\Sigma_i \, C_i \, B_\nu \, Q_{abs,i} / a$. Note that the calculated
values  of Q$_{abs}$/$a$ are independent of the grain size as long as
2$\pi |m| a/\lambda$ $<<$1 where $m$ is the complex refractive index.
Thus the derived scale factor C$_i$ as well as the estimated dust mass
(see \S4) are independent  of the radius of the dust.  The dust
compositions of the best fits are  summarized in Table 1. 

The strong 21 $\mu$m-peak dust is best fit by Mg proto-silicate,
amorphous  SiO$_2$ and FeO grains (with temperatures of 60-120 K) as
shown in Figure \ref{21umpeakspec}.  These provide a good match to the
21 $\mu$m feature.   ADM suggested that the 21 $\mu$m feature is best
fit by Mg proto-silicate  while D01 suggested it is best fit  by
SiO$_2$ instead.  We found, however, that SiO$_2$ produced a 21 $\mu$m 
feature that was too sharp.  We also fit the observations using
Mg$_2$SiO$_4$,  which exhibits a feature around 20 $\mu$m  and the
overall variation of absorption coefficients of Mg$_2$SiO$_4$ with
wavelength might be  similar to that of  Mg protosilicate \citep{dor80,
jag03}.  However, with Mg$_2$SiO$_4$, the fit is not as good as that
of  Mg protosilicate, not only at the 21-$\mu$m peak, but also  at
shorter (10-20 $\mu$m) and longer (70 $\mu$m) wavelengths. Thus, we
use  Mg protosilicate and SiO$_2$   as silicates to fit the 21
$\mu$m-peak dust feature.  The fit with Mg protosilicate, SiO$_2$ and
FeO is improved by adding aluminum oxide  (Al$_2$O$_3$, 83 K) and FeS
(150 K), where Al$_2$O$_3$ improved the overall continuum shape between
10-70 $\mu$m and FeS improved the continuum between 30-40 $\mu$m
(underneath the  lines of Si, S and Fe), as shown Figure
\ref{21umpeakspec}. The silicate composition is responsible for the 21
$\mu$m peak, suggesting that the dust forms around the inner-oxygen and
S-Si layers and is consistent with Ar being one of the oxygen burning
products.  We also include amorphous MgSiO$_3$  (480 K) and SiO$_2$
(300 K) to account for the emission  feature around the 9.8 $\mu$m. 
The composition of  the low temperature (40-90 K) dust component
necessary for  reproducing 70 $\mu$m is rather unclear.  Either 
Al$_2$O$_3$ (80 K) (Model A in Table 1) or  Fe (100 K) (Model B in
Table 1 and Figure ~\ref{modelBspec}) can fit equally well, as listed
in Table 1. We could use carbon instead of  Al$_2$O$_3$ or Fe,  but the
line and dust compositions suggest the emission is from inner O, S-Si 
layers, where carbon dust is not expected.  There are still residuals
in the  fit from the feature peaking at  21 $\mu$m (20-23 $\mu$m), and
an unknown dust feature at 11-12.5 $\mu$m (it is not a part of typical
PAH feature), as shown in Figure \ref{21umpeakspec}.  The former  may
be due to non-spherical grains or different sizes of grains.

The weak 21 $\mu$m continuum is  fit by FeO and  Mg$_2$SiO$_4$ or Mg
protosilicate (Models C and D in Table 1)  since  the curvature of the
continuum changes at 20-21 $\mu$m as shown in Figure
\ref{weak21umspec}.  To fit the rest of the spectrum, we use glassy
carbon dust and Al$_2$O$_3$ grains.  The glassy carbon grains (220 K)
can account for the  smooth curvature in the continuum between 8-14
$\mu$m. Carbon dust (80 K) and Al$_2$O$_3$ (100 K)  contribute  to the
continuum between 15-25 $\mu$m.  We could use Fe dust  instead, but we
suspect carbon dust because of the presence of relatively  strong Ne 
line emission with the weak 21 $\mu$m dust class.  Ne, Mg, and Al are
all  carbon burning products.   We cannot fit the spectrum replacing
carbon by Al$_2$O$_3$ with a single or two temperatures because
$Q_{abs}/a$ of Al$_2$O$_3$ has a shallow bump around 27 $\mu$m, thus
the fit requires three temperature components of  Al$_3$O$_2$ or a
combination of two  temperature components of  Al$_3$O$_2$ and  a
temperature component of carbon. The continuum between 33-40 $\mu$m
(underneath the  lines of Si, S and Fe) can be optimally fit by FeS
grains.

The 70 $\mu$m image shown in Figure \ref{images}e shows interior
emission  similar to the unshocked ejecta but that may also be due to
projected  circumstellar dust at the forward shock.  In order to fit
the featureless  spectrum out to 70 $\mu$m, we must first correct for
possible projected  circumstellar dust emission.  The exterior forward
shock emission is most  evident in the northern and northwestern
shell.  Taking the typical  brightness in the NW shell ($\sim$20
MJy~sr$^{-1}$), and assuming the  forward shock is a shell with 12\%
radial thickness,  the projected brightness  is less than 4-10\% of the
interior emission ($\sim$40 MJy~sr$^{-1}$ after background
subtraction).  We assume  that the remaining wide-spread interior 70
$\mu$m emission is from relatively  cold, unshocked ejecta.  Using the
``corrected'' 70 $\mu$m flux, the  featureless spectra are equally
reproduced by three models  (Models E, F, and G) in Table 1 and Figures
\ref{modelEspec} and \ref{flessspec}. All fits  include MgSiO$_3$, FeO
and Si, and either aluminum oxide, Fe, or a combination of the two are
required at long wavelength.  Carbon dust can also produce featureless
spectra at low temperature but we exclude this composition  because of
the lack of Ne (produced from carbon burning).  Aluminum oxide and  Fe
dust are far more likely to be associated with the unshocked ejecta 
because they result from O-burning and Si-burning, respectively and
the  unshocked ejecta exhibit Si, S, and O+Fe line emission.  However,
one of the  key challenges in SN ejecta dust is to understand
featureless dust such as  Fe, C, and aluminum oxide, and to link it to
the associated nucleosynthetic  products.

\section{Dust Mass}
We estimated the amount of freshly formed dust in Cas A based on our
dust  model fit to each of the representative 21 $\mu$m-peak, weak-21
$\mu$m, and featureless spectra (Fig. \ref{sixspec}). The dust mass of
$i$-grain type is given by: $$ M_{dust,i} = {F_\nu^i \, d^2 \over
{B_\nu (T_{d,i}) \, \kappa_i } }= {F_\nu^i d^2 \over {B_\nu(T_{d,i})}}
{ 4 \, \rho_i \, a \over {3 \,Q_{abs, i}}} $$ where $F_\nu^i$ is the
flux from $i$-grain species, $d$ is the distance,  $B_\nu$ is the
Planck function, $\rho_i$ is the bulk density, and $a$ is the dust
particle size.  By employing the scale factor $C_i$ and the dust
temperature  $T_{d,i}$ derived from the spectral fit, the total dust
mass is given by $M_{dust} = \sum_i \rho_{i} \Omega d^2 C_i/3$,  where
{\it $\Omega$} is the  solid angle of the source.  The total mass of
the 21 $\mu$m-peak dust  is then determined by summing the flux of all 
the pixels in the  21 $\mu$m-peak dust region (red region in Fig.
~\ref{images}f) and assuming  each pixel in this region has the same
dust composition as the spectrum in Fig.~\ref{21umpeakspec}.  We took
the same steps for the weak-21 $\mu$m  dust and the featureless dust.  

The estimated total masses for each type of dust using a distance of
3.4 kpc (Reed et al. 1995) are listed in Table 1.   Using the least
massive composition in Table 1 for each of the three dust classes
yields a total mass of \lowermass\ M$_\odot$ (the sum of masses from
Models A, D, and F). Using the most-massive composition for each of the
three dust classes yields a total mass of \uppermass\ M$_\odot$ (the
sum of masses from Models B, C, and E). The primary uncertainty in the
total dust mass between \lowermass\ and \uppermass\ M$_\odot$ is due to
the selection of the dust composition, in particular for the
featureless dust. 

We also extracted a global spectrum of Cas A , but excluding most of
the  exterior forward shock regions.  The spectrum is well fit with
the  combination of our three types of dust (including all
compositions  from Models A-G), as shown in Figure \ref{totalspec}.  We
used the dust composition of Models A-G as a guideline in fitting the
global spectrum, because  the dust features (which were noticeable in
representative spectra) were smeared out. Our goal in fitting the
global spectrum  is to confirm consistency between the mass derived
from global spectrum and that derived from  representative spectra
described above. The total estimated mass from the  global spectrum
fit  is  $\sim$0.028 M$_\odot$, being consistent to the mass determined
from the individual  fits to each dust class. The respective dust mass
for each grain composition is listed in Table 3.  The masses  of
MgSiO$_3$, SiO$_2$,  FeS and Si  are more than a factor of ten to
hundred smaller than  the predictions;   the predictions (N03 and TF)
also  have  the dust features at 9 $\mu$m for MgSiO$_3$, 21 $\mu$m for
SiO$_2$, and 30-40 $\mu$m for FeS stronger than the observed  spectra
if the dust mass is increased.  The carbon mass is also a factor of 10
lower than the predictions.   We  were not able to fit the data with as
much carbon dust mass as expected, even  if we use the maximum carbon
contribution allowed from the spectral fits.

\section{Discussion}

We find an estimated total freshly-formed dust mass of
\lowermass-\uppermass\  M$_\odot$  is required to produce the
mid-infrared continuum up to 70 $\mu$m.  The dust  mass we derive is
orders of magnitude higher than the two previous infrared estimates of 
3.5$\times$10$^{-3}$ M$_\odot$  and 7.7$\times$ 10$^{-5}$ M$_\odot$, 
which are derived by extrapolation  from 1.6$\times$10$^{-4}$ M$_\odot$
(D01) and 2.8$\times$ 10$^{-6}$ M$_\odot$ (ADM) for selected knots,
respectively. One of the primary reasons for our higher mass estimate
is that we  include fluxes up to 70 $\mu$m while the fits in D01 and
ADM accounted for dust emission only up to 30  and $\sim$40 $\mu$m,
respectively.   The cold dust (40-150 K) has much more mass than the
warmer ($>$150 K) dust.  In addition, our IRS mapping over nearly the
entire extent of Cas A with higher spatial and  spectral resolutions
provides more accurate measurements, while  D01 and  ADM covered only a
portion of the remnant.  In addition, ADM use only Mg protosilicate
dust; the absorption coefficient for Mg protosilicate is a few times
larger than those of other compositions.

Our dust mass estimate is also at least one order of magnitude higher
than the estimate of  3$\times$ 10$^{-3}$ M$_\odot$ by \cite{hin04}. 
They fitted MSX and \spitzer\ MIPS data with Mg protosilicate. Note
that they used only one composition.   They  derived a freshly
synthesized dust mass of 3$\times$10$^{-3}$ M$_\odot$  at a temperature
of  79-82 K and a smaller dust mass of 5$\times$10$^{-6}$ M$_\odot$  at
a higher temperature of 226-268 K, and they explained that the mass
estimate depends on the chosen dust temperature. As ADM mentioned, the
absorption coefficient for Mg protosilicate is a few times larger than
those of other compositions. Therefore, even including the
long-wavelength data, the estimated mass was small since only Mg
protosilicate was modeled. With the photometry in \cite{hin04}, one
could easily fit the data with only Mg protosilicate   and would not
need  additional grain compositions. However, with the accurate IRS
data, many dust features and the detailed continnum shape could not be
fit solely with the  Mg protosilicate.  Note that the continuum shapes
of weak 21 $\mu$m dust and ``featureless" dust are very different from
the shape of protosilicate absorption coefficient.  Therefore, it was
necessary to include many other compositions in order to reproduce the
observed IRS spectra.

It should be noted here that, in contrast with the previous works,  we
introduced Si and Fe--bearing materials such as Si, Fe, FeS and FeO.  We
explain why we included such dust in our model fitting as follows.
Firstly, we included Si and Fe dust because these elements are
significant outputs of nucleosynthesis; indeed \cite{woo95} show that
Si and Fe are primary products in the innermost layers of the ejecta. 
Secondly, we observed strong Si and Fe lines in the infrared and X-ray
spectra; strong Si lines were detected in the \spitzer\ spectra, as
shown in Figure \ref{sixspec} (also see D01), and the Fe line detection
at 17.9 $\mu$m is also shown in Figure \ref{weak21umspec}. (The Fe maps
at 17.9 $\mu$m and at 1.64  $\mu$m were presented in \cite{enn06} and
\cite{rho03}, respectively.)  Si and Fe lines from ejecta are also
bright in X-ray emission \citep{hwa00}.  Thirdly, dust such as Si, Fe,
FeO and FeS is predicted to form in the ejecta of Population III
supernovae (N03). TF and N03 predict Fe$_3$O$_4$ instead of FeO in the
uniformly mixed ejecta where the elemental composition is oxygen-rich, 
but the kind of iron-bearing grains in oxygen-rich layers of the ejecta
is still uncertain, partly because the surface energy of iron is very
sensitive to the concentration of impurities such as O and S (as was
discussed by \cite{koz88}), and partly because  the chemical reactions
at the condensation of Fe-bearing dust is not well understood.
Depending on the elemental composition and the physical conditions in
the ejecta, it is  possible that Fe, FeO and/or FeS form in the
oxygen-rich layers of Galactic SNe. The observations of Cas A favor FeO
dust over Fe$_3$O$_4$, in order to match the spectral shape of the 21
$\mu$m-peak dust and the weak-21 $\mu$m dust. This aspect should be
explored theoretically in comparison with the observations in the
future.

Our total mass estimate is also about  one order of magnitude higher
than the estimate of 6.9$\times$10$^{-3}$ M$_\odot$ by \cite{dwek87},
who used  IRAS fluxes (possibly confused by background cirrus)   and
assumed a silicate--type dust  as stellar or supernova condensates
being present in supernova cavity and heated up by the reverse shock.
Our estimated mass is much less than  1 M$_\odot$, which \cite{wil05}
suggested    may still be associated with the ejecta, after accounting
for results of  high-resolution CO observations.  Our estimated mass of
\lowermass\ to \uppermass\  M$_{\odot}$ is only derived for 
wavelengths up to 70 $\mu$m, so it is still possible that the total 
freshly-formed dust mass in Cas A is higher than our estimate because
there  may be colder dust present. Future longer-wavelength
observations with Herschel, SCUBA-2 and ALMA are required to determine
if this is the case. Also note that we did not include any mass from
fast moving knots projected into the same positions as the forward
shock, such as in the northeast and southwest jets, and the eastern
portions of the SNR outside the 21 $\mu$m-peak dust region (see Fig.
\ref{images}e),   because such dust could not be cleanly separated from
the interstellar/circumstellar dust.

We can use our dust mass estimate in conjunction with the models of N03
and  TF to understand the dust observed in the early universe.  If the
progenitor of Cas A was 15 M$_\odot$, our estimated dust mass
(\lowermass\--\uppermass\ M$_{\odot}$) is 7--18\% of  the 0.3
M$_{\odot}$ predicted by the models.   If the progenitor mass was 30
M$_\odot$, then the dust mass is 2--5\% of the 1.1 M$_{\odot}$
predicted by the models.  One reason   our dust mass is lower than
predicted by the models is that we cannot evaluate the  mass of very
cold dust residing in the remnant from the observered spectra up to 70
$\mu$m as described above, unless the predicted mass is overestimated. 
Another reason is that when and how much dust in the  remnant is swept
up by the reverse shock is highly dependent on the  thickness of the
hydrogen envelope at the time of explosion  and that the evolution and
destruction of dust grains  formed in SNe strongly depend not only on
their initial sizes   but also the density of ambient interstellar
medium (Nozawa et al. 2007). Dust formation occurs within a few hundred
days after the SN explosion (Kozasa et al. 1989; TF; N03). Without a
thick  hydrogen envelope, given an age for Cas A of $\sim$300 years, a
significant component of dust may  have already been destroyed if dust
grains formed in the ejecta  were populated by very small-sized grains;
otherwise,  it is possible that some grain types may be larger, which
would  increase the inferred mass.

We observed most of the dust compositions predicted by  SN Type II
models, and  the global ejecta composition is  consistent with the
unmixed-case  N03 model than mixed-case model; however, note that
different morphologies of Ar and Si maps imply that some degree of
mixing has occurred.     Our estimated dust mass with {\it Spitzer}
data is one order of  magnitude smaller than the predicted models of
dust formation in SNe ejecta  by N03 and TF, but one to two orders of
magnitude higher than the previous  estimations.   We now compare the
dust mass in high-redshift galaxies  with the observed dust mass of Cas
A based on the chemical evolution model of Morgan \& Edmunds (2001). 
By a redshift of 4, SNe have been injecting dust in galaxies for over 2
billion years and there is enough dust from SNe to explain the lower
limit on the dust masses ($\sim$7$\times$10$^7$ M$_\odot$) inferred in
submm galaxies and distant quasars \citep{chi94, isa02}. It should be
noted with the dust mass  per SN implied by our results for Cas A
alone, the interpretation of dust injection from SNe is limited, 
because the amount of dust built up over time is strongly dependent on
the initial mass function, stellar evolution models and star formation
rates \citep{mor03}, and destruction rates in supernova are believed to
be important at timescales greater than a few billion years. 
Additional infrared/submm observations of other young supernova
remnants and supernovae are crucial  to measure  physical processes of
dust formation in SNe including the dust size distribution, composition
and dependence on nucleosynthetic products and environment, and to
understand the dust in the early Universe in terms of dust injection
from SNe.

\section{Conclusion}
1. We presented \spitzer\ IRS mapping covering nearly the entire extent
of Cas A and examined if SNe are primary dust formation sites  that can
be used to explain the high quantity of dust observed in the early
Universe.  

2. The ejecta maps, show a remarkable similarity to the dust maps,
thereby confirming that dust formation occurs in the SN ejecta. 

3. The IRS spectra of Cas A show a few dust features such as an unique
21 $\mu$m peak in the continuum from Mg protosilicate, SiO$_2$, and
FeO.  We observed most of the dust compositions predicted by  SN Type
II dust models. However, the dust features in Cas A favour  Mg
protosilicate rather than Mg$_2$SiO$_4$, and FeO rather than
Fe$_3$O$_4$. The composition infers that the ejecta are  unmixed.  

4. Our total estimated dust mass with {\it Spitzer}  observations
ranging from 5.5 - 70 $\mu$m is \lowermass-\uppermass\ M$_{\odot}$, one
order of  magnitude smaller than the predicted models of dust formation
in SNe ejecta  by N03 and TF, but one or more orders of magnitude
higher than the previous  estimations.   The freshly formed dust mass
derived from Cas A is sufficient from SNe to explain the lower limit on
the dust masses in high redshift galaxies.

\acknowledgements
J. Rho thanks U. Hwang for helpful discussion of  X-ray emission of Cas
A.  This work is based on observations made with the \spitzer\ {\it
Space Telescope}, which is operated by the Jet Propulsion Laboratory,
California  Institute of Technology, under NASA contract 1407. Partial
support for this work was provided by NASA through an GO award issued
by JPL/Caltech. 

{}

\clearpage

\begin{deluxetable}{llccccccccccccclll}
\tabletypesize{\scriptsize}
\rotate
\tablewidth{0pt}
\tablecaption{Properties of Freshly Formed Dust in Cas A}
\label{catalog}
\tablehead{
\colhead{Dust Type} &
\colhead{Model} &
\colhead{Compositions\tablenotemark{a}} &
\colhead{Strong}  &
\colhead{Nucleosynthesis}   &
\colhead{ Mass }        \\
\colhead{(spectrum in Fig. \ref{sixspec})} &
\colhead{} &
\colhead{} &
\colhead{ Lines}  &
\colhead{Layers}   &
\colhead{(M$_{\odot}$)} \\
}
\startdata
21$\mu$m-peak (a)  &A& {\bf Mg protosilicate}, {\bf  MgSiO$_3$}, SiO$_2$,
FeO, FeS, Si,
{\it Al$_2$O$_3$}   & Ar & Inner-O, S-Si  &  0.0030  \\
21$\mu$m-peak  &B& {\bf Mg protosilicate}, {\bf  MgSiO$_3$}, FeO, SiO$_2$, FeO, FeS, Si,
{\it  Fe}   & Ar & Inner-O, S-Si  &  0.0120   \\
weak-21$\mu$m (b) & C  & {\bf C-glass}, {\bf FeO}, Al$_2$O$_3$,  Si, {\it Mg$_2$SiO$_4$} 
& Ne, Si, Ar (S, O+Fe) &  C-burning & 0.0180  \\
weak-21$\mu$m &  D & {\bf C-glass}, {\bf FeO}, Al$_2$O$_3$, Si, FeS, 
{\it Mg protosilicate} 
& Ne, Si, Ar (S, O+Fe) & C-burning & 0.0157  \\
Featureless (d)  &  E  & {\bf MgSiO$_3$},  {\bf Si}, FeS, {\it Fe,  Mg$_2$SiO$_4$} & Si, S, (O+Fe)  & O, Al burning (Fe-Si-S)  & 0.0245  \\
Featureless &  F  & {\bf MgSiO$_3$}, {\bf Si}, FeS,  {\it Fe, Al$_2$O$_3$} & Si, S, (O+Fe)  & O, Al burning (Fe-Si-S)  & 0.0171\\
Featureless &  G  & {\bf MgSiO$_3$}, {\bf Si}, FeS,  {\it Al$_2$O$_3$, Mg$_2$SiO$_4$}  & Si,
  S, (O+Fe)  & O, Al burning (Fe-Si-S)  & 0.0009\\
\enddata
\tablenotetext{a}{Compositions in the best fit, where 
a few primary compositions are written in bold, and
alternative dust compositions are in italics.
}
\end{deluxetable}

\begin{deluxetable}{lllllllllllllll}
\tabletypesize{\scriptsize}
\rotate
\tablewidth{0pt}
\tablecaption{Mass of dust with each composition}
\label{dustmasstab}
\tablehead{
\colhead{Composition} &
\colhead{Model A\tablenotemark{a}} &
\colhead{ Model B} &
\colhead{Model C}  &
\colhead{Model D}   &
\colhead{Model E }  &
\colhead{Model F} &
\colhead{Model G} & \\
}
\startdata
     Al$_2$O$_3$  &  6.66E-05 (083) &  0.00E+00 (000) &  5.13E-05 (105) &  1.03E-04 (100) &  0.00E+00 (000) &  8.13E-04 (050) &  6.50E-04 (060) &\\
         C glass  &  0.00E+00 (000) &  0.00E+00 (000) &  2.08E-03 (80/180) &  1.07E-03 (80/220) &  0.00E+00 (000) &  0.00E+00 (000) &  0.00E+00 (000) &\\
       MgSiO$_3$  &  1.19E-08 (480) &  1.19E-08 (480) &  0.00E+00 (000) &  0.00E+00 (000) &  2.55E-05 (110) &  3.19E-05 (110) &  2.55E-05 (110) &\\
   Mg$_2$SiO$_4$  &  0.00E+00 (000) &  0.00E+00 (000) &  7.89E-05 (120) &  0.00E+00 (000) &  1.72E-06 (130) &  0.00E+00 (000) &  3.00E-06 (130) &\\
 Mg protosilicate  &  5.00E-05 (120) &  4.67E-05 (120) &  0.00E+00 (000) &  3.77E-05 (120) &  0.00E+00 (000) &  0.00E+00 (000) &  0.00E+00 (000) &\\
         SiO$_2$  &  2.23E-03 (060/300) &  1.40E-03 (065/300) &  0.00E+00 (000) &  0.00E+00 (000) &  0.00E+00 (000) &  0.00E+00 (000) &  0.00E+00 (000) &\\
              Si  &  4.34E-04 (096) &  4.34E-04 (100) &  1.63E-03 (090) &  8.17E-03 (080) &  9.32E-04 (090) &  1.24E-04 (120) &  6.21E-05 (120) &\\
              Fe  &  0.00E+00 (075) &  9.82E-03 (110) &  0.00E+00 (000) &  0.00E+00 (000) &  2.16E-02 (95/135) &  1.36E-02 (100/150) &  0.00E+00 (000) &\\
             FeO  &  1.13E-04 (105) &  2.11E-04 (095) &  1.39E-02 (060) &  5.97E-03 (065) &  0.00E+00 (000) &  0.00E+00 (000) &  0.00E+00 (000) &\\
             FeS  &  1.20E-04 (150) &  2.11E-04 (150) &  0.00E+00 (000) &  3.40E-04 (120) &  1.94E-03 (055) &  2.59E-03 (055) &  1.29E-04 (100) &\\
\enddata
\tablenotetext{a}{The numbers in parentheses are
dust temperatures, and two numbers indicate two temperatures.}
\end{deluxetable}

\begin{deluxetable}{llllllllll}
\tabletypesize{\scriptsize}
\tablewidth{0pt}
\tablecaption{Mass of dust with each composition}
\label{dustmasstab}
\tablehead{
\colhead{Composition} &
\colhead{Predicted ($M_\odot$)} &
\colhead{Model (A+D+G)\tablenotemark{a}} &
\colhead{Model (B+C+E)\tablenotemark{b}}  &
\colhead{Mass (global)\tablenotemark{c}}  
}
\startdata
Al$_2$O$_3$      & 2.40E-04 $\sim$ 9.00E-03    &8.20E-04 &5.1300E-05 &1.22E-04 (105) \\
carbon           & 7.00E-02 $\sim$ 3.00E-01    &1.07E-03 &2.0767E-03 &2.04E-03 (070/265) \\
MgSiO$_3$        & 2.00E-03 $\sim$ 7.00E-3     &2.55E-05 &2.5500E-05 &1.65E-04 (110) \\
Mg$_2$SiO$_4$    & 3.70E-02 $\sim$ 4.40E-1     &3.00E-06 &8.0620E-05 &3.21E-05 (120) \\
Mg protosilicate & \tablenotemark{d}none                    &8.77E-05 &4.6710E-05 &6.70E-05 (110) \\
SiO$_2$          & 2.50E-02 $\sim$ 1.400E-01   &2.23E-03 &1.3964E-03 &1.35E-03 (065) \\
Si               & 7.00E-02 $\sim$ 3.00E-01    &8.66E-03 &2.9989E-03 &4.42E-03 (080) \\
Fe               & 2.00E-02 $\sim$ 4.00E-02    &0.00E+00 &3.1459E-02 &1.03E-02 (090) \\
FeO              & \tablenotemark{d}none                    &6.08E-03 &1.4136E-02 &6.23E-03 (070) \\
FeS              & 4.00E-02 $\sim$ 1.10E-01    &5.90E-04 &2.1501E-03 &2.90E-03 (090) \\
\enddata
\tablenotetext{a}{The sum of the masses using the least massive composition among models.}
\tablenotetext{b}{The sum of the masses  using the most-massive compositions.}
\tablenotetext{c}{The mass  using the global spectrum.}
\tablenotetext{d}{see the text for details.}
\end{deluxetable}

\clearpage

\clearpage

\begin{figure}
\caption{ (a) 21 $\mu$m dust map; a continuum map of 19-23 $\mu$m
subtracted by the baselines of neighboring wavelengths.  This dust map
is remarkably similar to the [Ar II] map (b) the resolutions are
convolved to match to each other).  The image is centered at R.A.\
$23^{\rm h} 23^{\rm m} 25.86^{\rm s}$ and Dec.\ $+58^\circ$49$^{\prime}
14^{\prime \prime}$ (J2000), and covers an 7.87$'$ by 5$'$ field
of view. (c) A combination of [O~IV] and [Fe~II] line map at 26 $\mu$m.
(d)  [Si~II] (34.8 $\mu$m) map. (e) MIPS 70 $\mu$m map from \cite{hin04}.
The locations of the forward and reverse shock boundaries are marked as 
ellipses where we adjusted for elongation from \cite{got01} using
long-exposure {\it Chandra} archival data \citep{hwa04}.
(f) Distributions of three major groups of dust types :
21$\mu$m-peak dust regions are in red (spectrum (a)), weak-21$\mu$m
dust regions are in green (spectrum (b)), and featureless-dust
regions are in blue ((d) spectrum in Fig.~\ref{sixspec}).
{\it (Fig. 2 is a jpeg file)} }
\label{images}
\end{figure}

\begin{figure}
\plotone{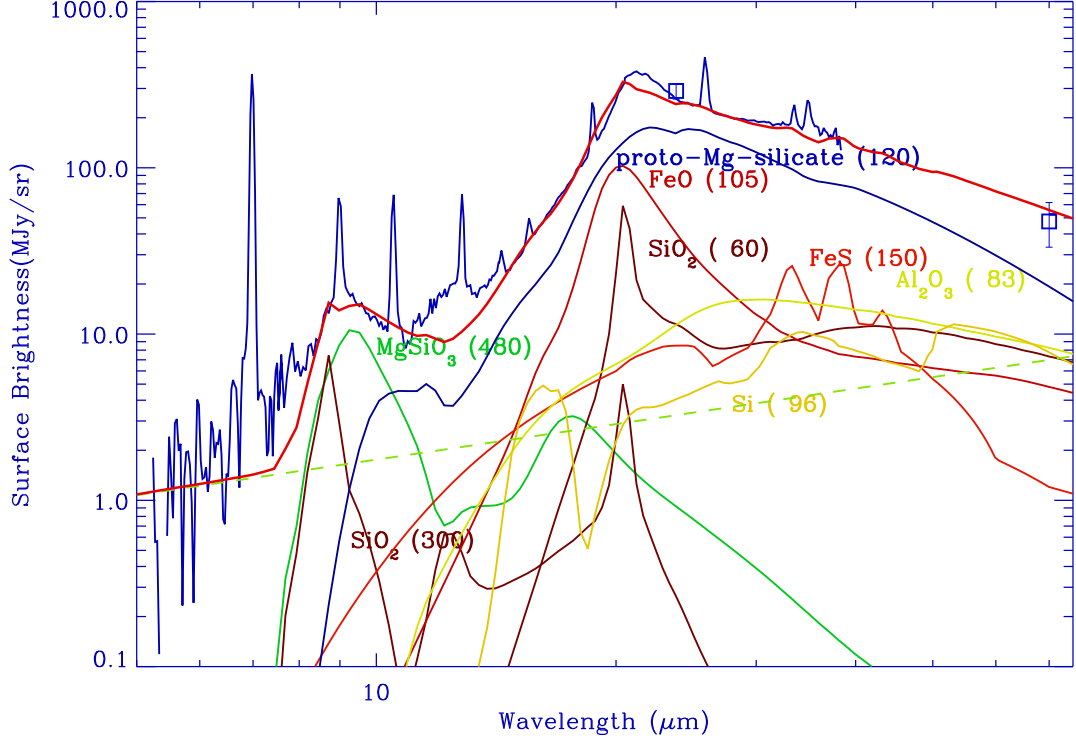}
\caption{21 $\mu$m-peak dust spectrum  superposed on the dust fit  of
Model A: A {\it Spitzer} IRS spectrum towards a bright part of the
northern shell fitted with dust compositions of Mg proto-silicate,
MgSiO$_3$,  SiO$_2$, FeO,  and Al$_2$O$_3$. The compositions suggest
that the dust forms around inner-oxygen and S-Si layers. The data and
the total fit are shown in blue and thick red lines, respectively, and
MIPS fluxes are marked with squares. The dust temperatures are shown
in  parentheses, and the dotted lines are from the second temperature
components.  Synchrotron continuum contribution (green dashed line) is
estimated  based on the radio fluxes and IRAC 3.6$\mu$m image.}
\label{21umpeakspec}
\end{figure}

\begin{figure}
\plotone{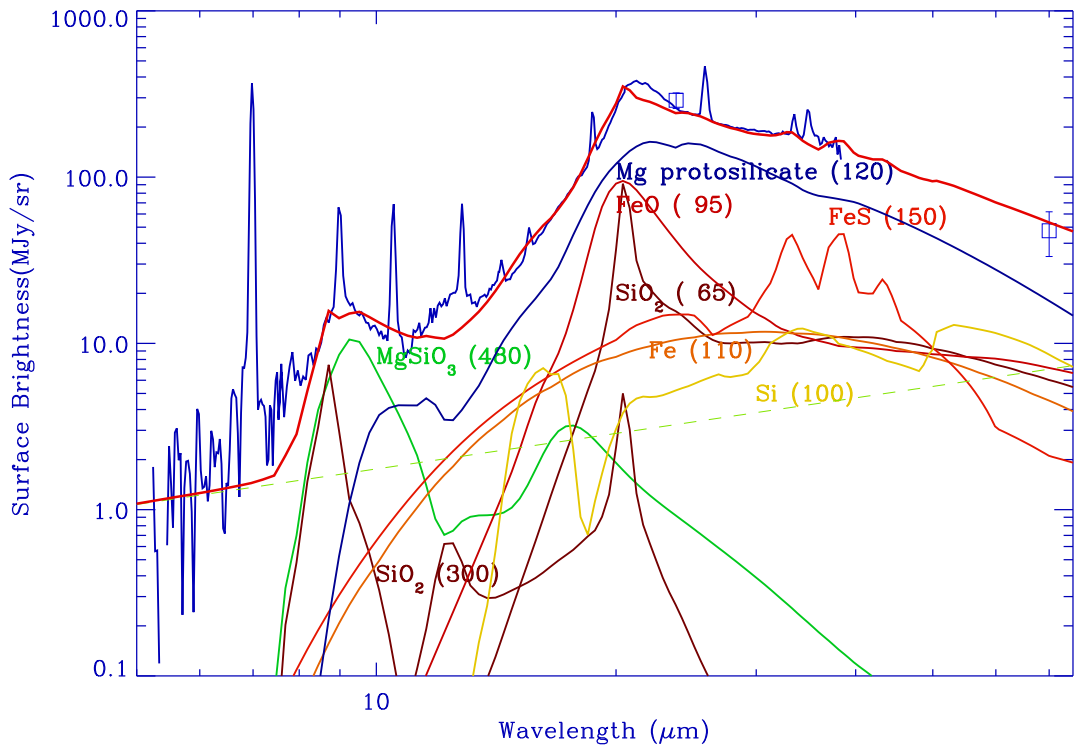}
\caption{21 $\mu$m-peak dust superposed on the dust fit of Model B. }
\label{modelBspec}
\end{figure}

\begin{figure}
\plotone{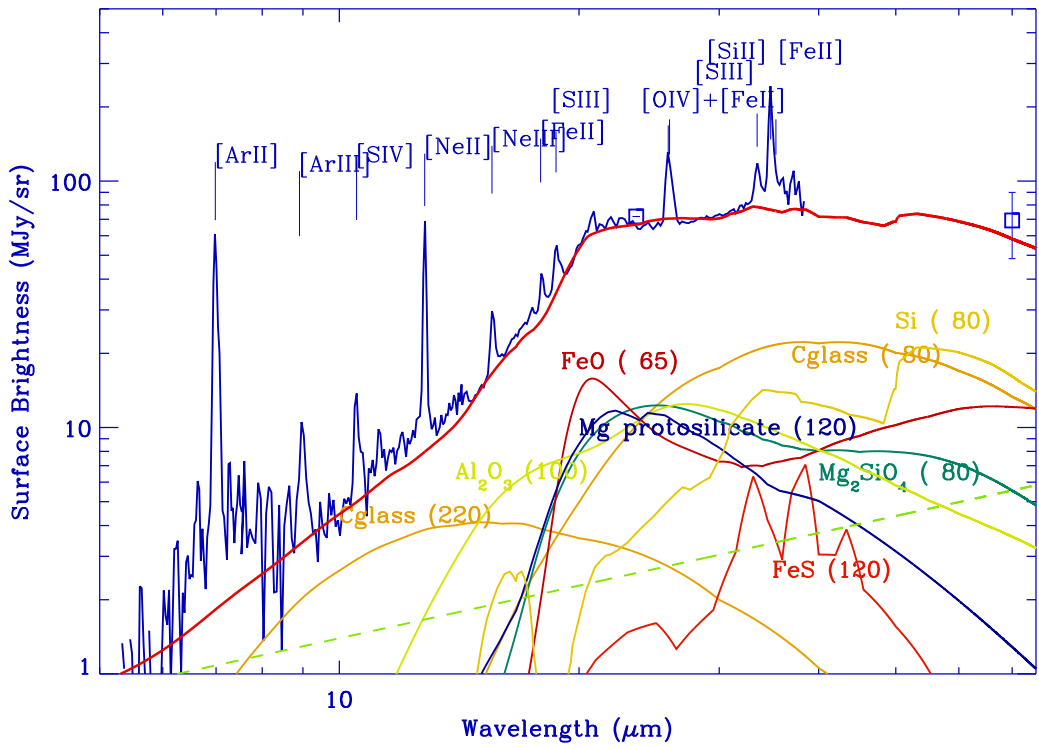}
\caption{Weak-21 $\mu$m dust superposed on the dust fit of Model D: A
second type of dust continuum in Cas A.  The distribution of this type
of dust is shown in Fig.~\ref{images}f, in  green. }
\label{weak21umspec}
\end{figure}

\clearpage

\begin{figure}
\plotone{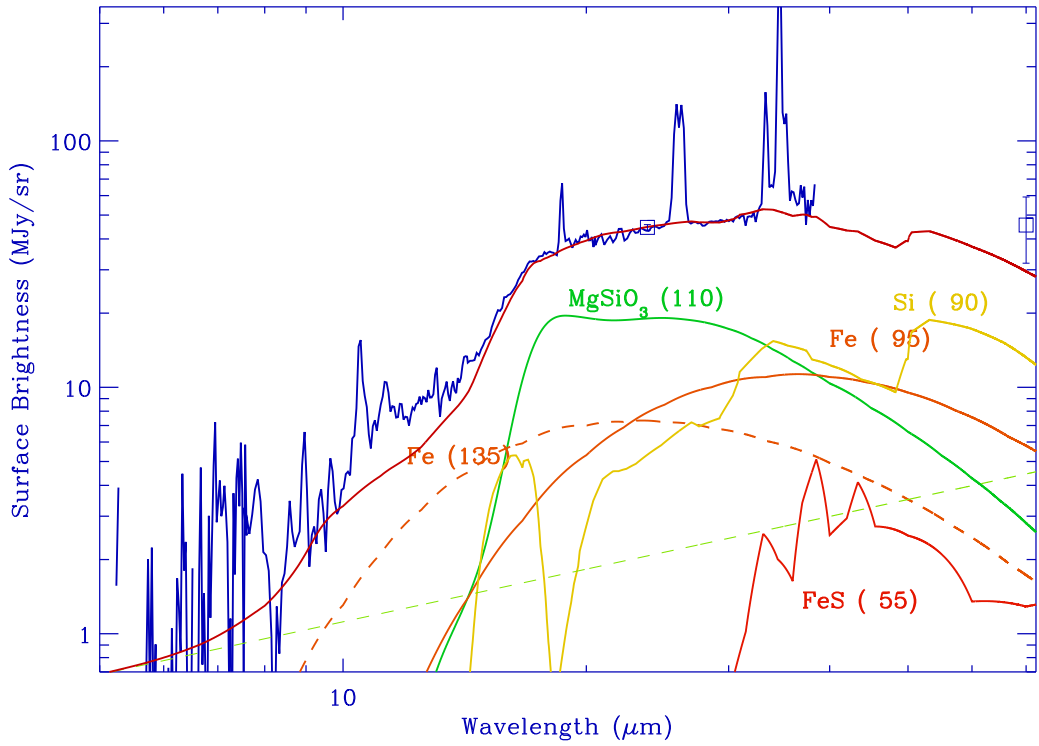}
\caption{
Featureless Dust Spectrum:
the continuum can be fit with MgSiO$_3$ and Fe (Model E).
The featureless spectra accompanies with S, Si, and O/Fe lines.
 The green dashed line is predicted synchrotron
emission model.}
\label{modelEspec}
\end{figure}

\begin{figure}
\plotone{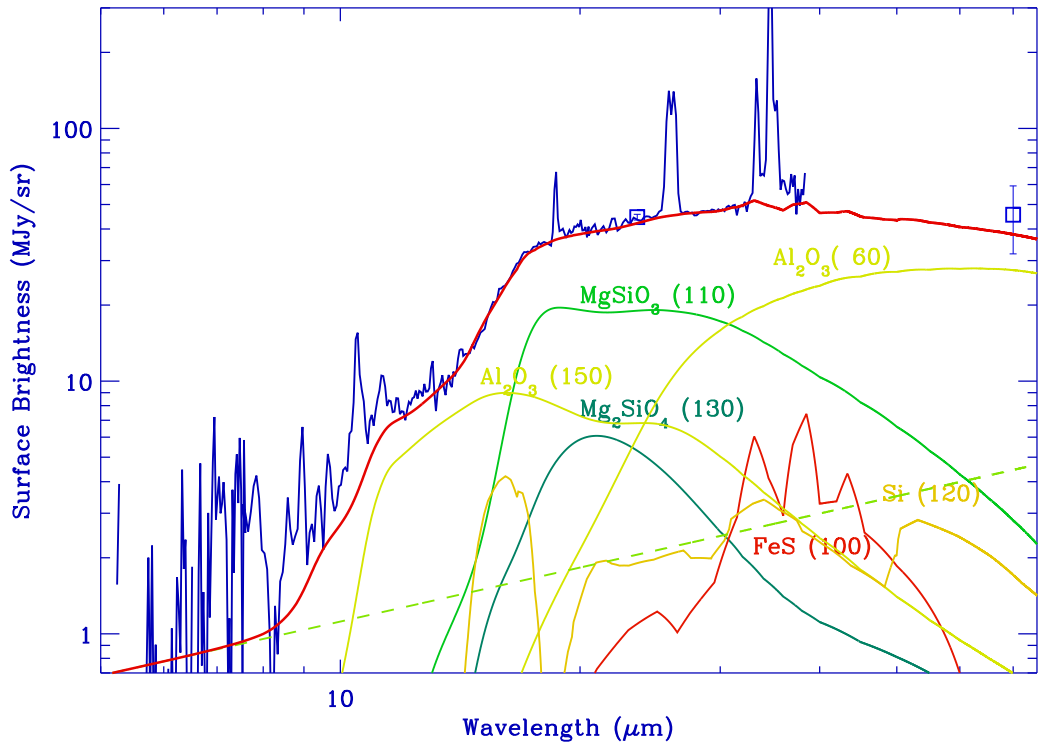}
\caption{ 
Featureless Dust Spectrum:
The continuum can be fit with Al$_2$O$_3$, 
and MgSiO$_3$ (Model G).
}
\label{flessspec}
\end{figure}

\begin{figure}
\plotone{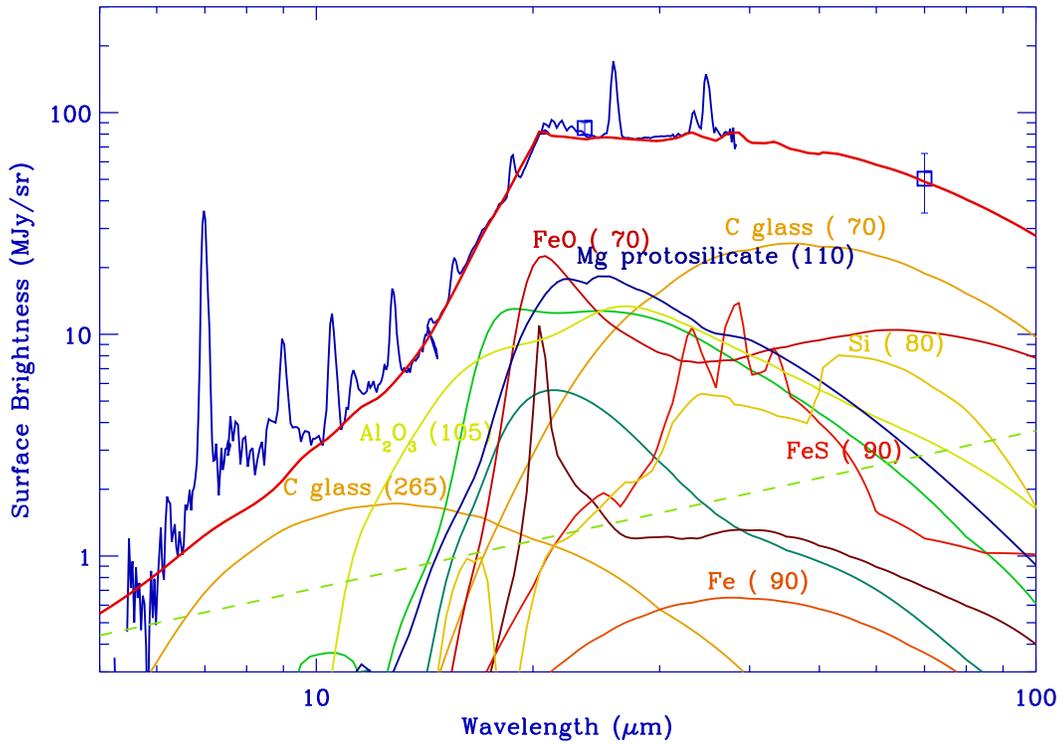}
\caption{Total spectrum of Cas A and the dust fit of grains with a few
primary contributions of grain models. 
The dark green and brown lines are MgSiO$_3$ (100 K) and SiO$_2$ (65K), respectively.  
 The green dashed line is predicted synchrotron
emission model.}
\label{totalspec}
\end{figure}

\end{document}